\documentstyle[epsfig]{europhys}

\newif\ifboo \boofalse

\begin{document}
\euro{}{}{}
\Date{}
\title{Microphase  Separation  and   modulated phases  in    a Coulomb
frustrated  Ising   ferromagnet}  \author{P.  Viot  and   G.   Tarjus}
\institute   {  Laboratoire  de    Physique  Th\'eorique  des Liquides
Universit\'e Pierre et Marie Curie 4, place  Jussieu 75252 Paris Cedex
05 France} 

\rec{}{}

\pacs{ \Pacs{05}{50$+$q}{Lattice theory
and statistics; Ising problems}
 \Pacs{05}{70.Fh}{Phase   transition:
general aspects} 
  \Pacs{64}{60.Cn}{Order-disorder    and  statistical
mechanics of model systems} }

\maketitle

\begin{abstract}
We  study a 3-dimensional Ising model  in  which the tendency to order
due  to  short-range   ferromagnetic  interactions  is  frustrated  by
competing long-range (Coulombic) interactions. Complete  ferromagnetic
ordering is impossible for    any  nonzero value of  the   frustration
parameter,  but the system displays a  variety of phases characterized
by periodically  modulated structures.   We have   performed extensive
Monte-Carlo simulations   which   provide strong   evidence that   the
microphase separation  transition  between paramagnetic and  modulated
phases is  a  fluctuation-induced  first-order transition.  Additional
transitions to various commensurate phases may also occur when further
lowering the temperature. 
\end{abstract}

In a variety  of systems a tendency   to order induced  by short-range
interactions is frustrated by long-range competing effects. In diblock
copolymer melts,  cross-linked polymer mixtures  and  interpenetrating
networks,  short-range forces between mutually incompatible components
would drive a phase transition  at  low enough temperature, but  total
segregation is forbidden by   constraints  due to covalent  bonds   or
crosslinks.   Instead, a microphase  separation transition  may occur,
where the system forms  phases with periodically  modulated structures
such          as         lamellar,        hexagonal,    or       cubic
phases\cite{leibler,bates,fried}.     Similarly,  self-assembly     in
water-oil-surfactant mixtures results from the competition between the
tendency   of oil and water  to  phase separate and the stoichiometric
constraints due  to the presence of the  surfactant molecules that act
as  the  electroneutrality    condition  in  a    system of    charged
particles\cite{WCS,carraro}.  In    supercooled liquids,  the dramatic
slowing down  of the structural  relaxation which  leads to  the glass
transition has been  ascribed  to the presence  of frustration-limited
domains  whose   formation comes from  the  inability  of  the locally
preferred  arrangement of the molecules  in  the liquid to tile  space
periodically\cite{KKZNT}.    Others examples include frustrated  phase
transition in   doped   antiferromagnets\cite{EK}  and   metal-ammonia
solutions\cite{carraro},  or   else  pattern  formation   in   various
nonequilibrium systems\cite{SH}. 

The physics of the above mentioned situations appears to be reasonably
well  described by   lattice  or   continuum  models  in   which  both
short-range ordering and long-range Coulombic frustrating interactions
are present.   In this letter, we  consider a Coulomb frustrated Ising
ferromagnet whose Hamiltonian is given by

\begin {equation}\label{hamil}
H=-J\sum_{<i,j>}S_iS_j+\frac{Q}{2}\sum_{i\neq j}\frac{S_iS_j}{r_{ij}}, 
\end {equation}
where $J,Q>0$, $S_i=\pm   1$, $<i,j>$ indicates  a   sum restricted to
nearest neighbors, and $r_{ij}$ is  the distance between the sites $i$
and $j$ on a three-dimensional cubic  lattice.  The two main questions
that  we address here by combining  analytical results and Monte-Carlo
simulations  are  the   following:  (i) what   is  the  nature of  the
microphase  separation   transition from   the   paramagnetic   to the
``modulated'' phases and  how  does it vary with  varying frustration,
i.e.  varying ratio  $Q/J$?  (ii) What is the  nature of the modulated
phases? 

The mean-field approximation to the Coulomb frustrated model is easily
derived\cite{GTV}.  We introduce  $\tau_{{\bf k}} (Q)$  as the Fourier
transform of the  full  pair interaction potential  in Eq.\ref{hamil};
e.g., for small  wave vectors and a  $3-d$  cubic lattice, $\tau_{{\bf
k}} (Q)$ is approximatively given by 
\begin{equation}\label{tauk}
\tau_{{\bf     k}} (Q)\simeq-6+\Delta       ({\bf k})  
+\frac{4\pi    Q}{\Delta   ({\bf   k})}\left(1+    b\Delta  ({\bf k})
+c\frac{\sum_{\alpha=x,y,z}\Delta^2(k_\alpha)}{\Delta({\bf
k})}+\cdots\right), 
\end{equation}
where       $\Delta({\bf       k})=\sum_{\alpha=x,y,z}\Delta(k_\alpha)
=2\sum_{\alpha=x,y,z}(1-\cos(k_\alpha))$  is  the Fourier transform of
the lattice Laplacian, $b$ and $c$ are constants, and $J$ has been set
equal  to 1.  For  $Q\neq 0$, $\tau_{{\bf  k}}  (Q)$ attains its
minimum, $\tau_{m} (Q)={\rm Min}_{\bf k}\{{\tau_{{\bf k}} (Q)}\}$, for
a set of nonzero values of the wave vector, $\{ {\bf k}_m (Q)\}$.  The
mean-field  treatment  predicts   the  occurrence   with    decreasing
temperature of  a second-order phase  transition at  $T_c(Q)=-\tau_{m}
(Q)$; $T_c(Q)$ goes continuously  to $T^0_c$, the critical temperature
of the unfrustrated  ferromagnet,  as $T^0_c-T_c(Q)\sim Q^{1/2}$  when
$Q\rightarrow 0$.  At the transition, the ordering is characterized by
a nonzero wave vector, corresponding to one of  the ${\bf k}_m (Q)$'s,
which for  small enough values  of $Q$ is  zero in two  directions and
behaves as $Q^{1/4}$ in the third one.   The mean spherical version of
Eq.(\ref {hamil}),  in which  the spins  are taken  to be real numbers
with  the global constraint that their  mean square  value is equal to
one, leads to  a  quite different behavior\cite{GTV,CEKNT}.  The  most
significant feature is the existence of  an ``avoided'' critical point
at $T^0_c$\cite{CEKNT}.  The microphase separation transition is again
of  second-order and the ordering  wave vector is one  of the $\{ {\bf
k}_m (Q)\}$'s; but the critical temperature is  now given by $T^{-1}_c
(Q)=\int  d^3k/(2\pi)^3 (\tau_{{\bf k}}  (Q) -\tau_{m} (Q))$, and when
$Q\rightarrow O+$ it goes  to a  temperature which  is much less  than
$T^0_c$.   Actually,  within   the    mean spherical treatment,    the
transition  is  very  sensitive   to the  form    of  the long   range
interaction:  whereas a $1/r$     interaction leads to $T_c(Q)>0$,   a
slightly  different   form obtained  by   retaining only   the leading
$1/\Delta({\bf  k})$ term in Eq.(\ref{tauk})   drives $T_c(Q)$ to zero
for $Q<16$\cite{CEKNT}.  This  latter  feature results from  the  fact
that the minimizing wave vectors ${\bf  k}_m (Q)$ form a $2-d$ surface
in ${\bf k}-$space (they form a finite set for the $1/r$ interaction).
A  similar   feature  is  encountered  in   the Landau-Ginzburg-Wilson
Hamiltonian that describes symmetric diblock copolymers\cite{leibler}.
It  has   been argued  by   Brazovskii\cite{B},  on the   basis   of a
self-consistent  Hartree approximation,  that  such systems  display a
fluctuation-induced first-order  transition  that  takes  place at   a
nonzero  temperature $T_0(Q)$,  with   $T^0_c-T_0(Q)\sim Q^{1/4}$ when
$Q\rightarrow O+$.  Additional  support  to this prediction  has  been
given by renormalization group analyzes\cite{MH}.

In order to check the relevance  of these various predictions, we have
performed extensive  Monte-Carlo  simulations of  the  $3-$dimensional
Coulomb frustrated Ising ferromagnet.  Our results provide, we believe
for   the first time,   direct    numerical evidence  supporting   the
Brazovskii  scenario of  a fluctuation-induced first-order  transition
for such a model.   The simulations have been done  for many values of
the frustration parameter $Q$.  Long-range  interactions lead to  very
time-consuming runs and we have considered lattices of $L^3$ sites for
$L$  ranging   from $4$ to   $24$   with periodic  boundary conditions
(anisotropic  lattices of various sizes  up  to $12\times 12\times 30$
have been  also  used).  To  treat properly  the  Coulomb interactions
Ewald sums have  been used; the  site-site  pair terms are  calculated
once for all at the  beginning of the  run and are  stored in an array
for  the entire run: therefore,  a large  number of reciprocal vectors
can be  included in the  Ewald sum to ensure  a very good accuracy for
the calculation of  the  Coulomb potential.   The  constraint of  zero
total   magnetization has been    enforced  by always  considering the
simultaneous  flip of a   pair of up   and down spins  (note that  the
presence   of  long-range   interactions    and  the  relatively  fast
equilibration    for  system sizes studied  reduce    the advantage of
considering cluster algorithms).  Each run consists then of $2-6\times
10^4$ Monte-Carlo Sweeps   (MCS)  per lattice site for   equilibration
followed  by  $6\times 10^4-10^5$  MCS   to calculate the  equilibrium
quantities.  In order to verify  that the equilibrium state is  indeed
reached  and  that  the   results do   not depend  upon   the  initial
configuration,  each   run is   performed  by   starting both  from  a
high-temperature configuration and from the ground state.

In  Fig.   1,  we display the   microphase separation  transition from
paramagnetic       to  modulated      phases     in    the    (scaled)
temperature-frustration diagram  for a dozen of values  of  $Q$.  Also
shown on  the   same diagram are  the   mean-field  and mean-spherical
predictions.  Transitions associated with very small values of $Q$ are
of course  inaccessible to simulations in  finite-size boxes (at least
two lamellae characterized by opposite  magnetizations must be present
in   the simulation box  to   satisfy  the  constraint  of zero  total
magnetization,   which  limits the width     of  the lamellae  and  by
consequence the smallness  of  the frustration); however, the  results
are compatible with a power-law behavior,
\begin{equation}
T^0_c-T_0(Q)\sim Q^\kappa
\end{equation}
with $\kappa\simeq 0.25-0.35$,    and   the period of  the    lamellar
structures is    in good agreement    with  $2\pi/|{\bf k}_m(Q)|$.  As
stressed before, the    transition  is first-order.  To  locate    the
transition and determine its order, we have monitored as a function of
both decreasing and   increasing temperature  the  energy  histograms,
which provides the total  energy per spin $E$  of the system, the heat
capacity $C_v=<(E-<E>)^2>$  and  the Binder  cumulants;  we  have also
studied  the (potential) order  parameter $|M_{\bf k}|=|<S_{\bf k}>|$,
where $S_{\bf k}$ is the Fourier transform  of the spin variable along
the three directions of the lattice.  Results for $Q=0.4$ are shown in
Fig.  2.  A hysteresis  loop  between heating and cooling
runs is observed in $E$ versus $T$ (Fig.  2a), 
and the distribution of the energy
histograms  around  the   transition temperature  $T_0$  exhibits    a
forbidden region for the energy (Fig.  2b); they both indicate a
first-order  transition.   Also supporting   this  conclusion  is  the
apparent  jump    in  the  order  parameter    and  the  strong system
size-dependence of  the  heat  capacity (e.g.   ${\rm Max}_TC_v(T)\sim
L^p$ with  $p>1$).  Similar behavior is  observed for the other values
of $Q$.     However,  the transition  becomes   more and   more weakly
first-order as the frustration decreases.

\begin{figure}
\begin{center}
\resizebox{5.2cm}{!}{\includegraphics{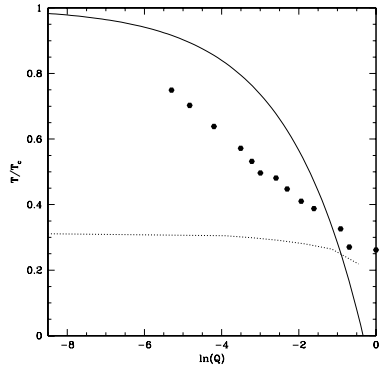}}
\end{center}
\caption[99]{Location of  the microphase separation  transition in the
$T/T^0_c-ln(Q)$  diagram.  (Recall that  $J=1$.)   The dots denote the
Monte-Carlo results ($T^0_c\simeq 4.51$)  The full line corresponds to
the mean-field prediction  ($T^0_c=6$) and   the  dotted line to   the
mean-spherical prediction ($T^0_c=3.91...$).}
\end{figure}

To provide a more complete investigation of the phase diagram, we have
also studied analytically and  numerically the $T=0$ properties of the
system.   In doing so,  we  have assumed,  as  indeed observed in  our
Monte-Carlo simulations  and  also in several  related two-dimensional
models\cite{LEFK,mcisaac} and   in  the three-dimensional  Ising model
with  dipolar competing   interactions\cite{binder},  that the  ground
states   consist   in  periodically    modulated   structures.    When
$Q\rightarrow  +\infty$   the ground  state  is an  anti-ferromagnetic
(N\'eel)  state $(1\times 1\times   1)$  and it  is stable  for $Q\geq
15.3$.    For  $15.3\geq Q\geq  14.6$,   the system  passes  though  a
succession   of   ``anisotropic  checkerboard''  phases  of   the form
$(1\times  1\times n)$  with  $n$ a  positive integer.  A  ``tubular''
phase $(1\times 1\times  \infty)$ is stable  for $14.6\geq Q\geq 6.2$,
and  in a narrow range   of $Q$ ($6.2\geq Q\geq   5.2$), one obtains a
second  infinite sequence of $(1\times  n\times  \infty)$ phases.  For
smaller values of  $Q$, the ground  state is found  by lamellar phases
whose width increases with decreasing  $Q$.  We have checked that more
complicated  phases  involving  mixtures  of  lamellae  with different
widths are not stable at zero  temperature.  When $Q\rightarrow 0$ and
for the purpose of calculating   the energy, the Fourier transform  of
the   Coulomb potential can be    well approximated by $1/\Delta ({\bf
k})$.  For a lamellar phase of period  $2m$ in the $z-$ direction, the
only nonzero values of $|S_{\bf   k}|$ are for ${\bf k}=(0,0k_z)$  and
$|S_{k_z}|=1/(m\sin   (k_z/2))$     with      $k_z=\pi(2n+1)/m$    and
$n=0,...,m-1$.    Therefore, after some   calculation, one  obtains an
explicit expression for the corresponding energy per spin,
\begin{equation}
E(m)=-J(3-\frac{2}{m})+Q\left(\frac{\pi}{6}m^2+\frac{\pi}{3}\right), 
\end{equation}
showing  that the lamellar width  $m$  goes as $(\pi Q/6)^{1/3}$  when
$Q\rightarrow  0$.  This dependence is   analogous to that derived for
symmetric diblock  copolymer     systems  in the  so-called     strong
segregation   limit\cite{leibler,bates}, whereas  the  behavior at the
transition  line to the paramagnetic  state  rather corresponds to the
weak-segregation limit result\cite{leibler,bates}.

At nonzero  temperatures, one may  expect that the system  undergoes a
series  of transitions to  various  commensurate, and at  temperatures
close  to $T_0(Q)$   incommensurate,  phases just  like in  the  ANNNI
model\cite{selke}.      Such a behavior  is indeed     observed in the
mean-field  theory\cite{GTV}, but  a  sequence of  modulated phases is
hard to detect in Monte-Carlo  simulations of finite-size systems.  We
have focused  our study on $Q=0.144$ for  which the ground state  is a
lamellar  structure   of half-period   $2$ and  the  ordering   at the
microphase  separation  transition is  characterized  by  a modulation
half-period  close to  2.5.  (Other  modulations  could  also occur  at
intermediate temperatures.)  To  observe at  least these two  types of
order in a simulation where only restricted values of the wave vectors
are allowed, we have used anisotropic  lattices.  For instance, with a
$12\times 12\times 24$ lattice, phases  with modulation of half-period
of $2.4$ are allowed in the z-direction.   Figure 3 illustrates at the
level  of  the  order parameter  $|M_{\bf   k}|$ the  passage from the
low-temperature  state  formed  by periodic    lamellae of  width  $2$
$(k_z=\pi/2)$ to a mixed or  incommensurate phase characterized by  an
average  half-period of  $2.4$  $(k_z=5\pi/12)$   for  $1.85\geq T\geq
1.75$.  For $T\geq 1.85$, the system  is disordered and all components
of $|M_{\bf   k}| $ are zero.   Corresponding  spin configurations are
displayed in Fig.4.

\begin{figure}\begin{center}
 \resizebox{10.6cm}{4.8cm}{\includegraphics{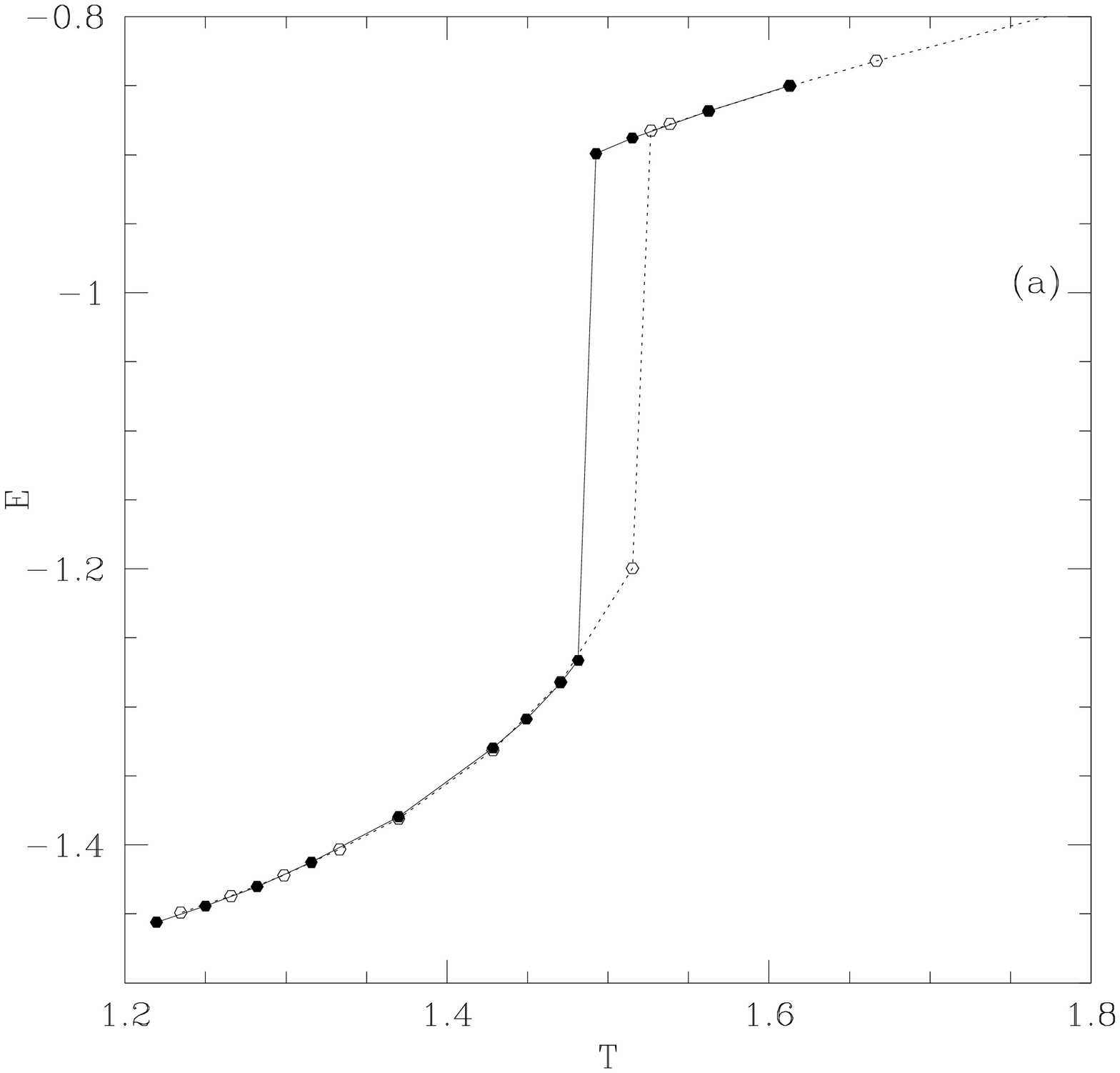}
\includegraphics{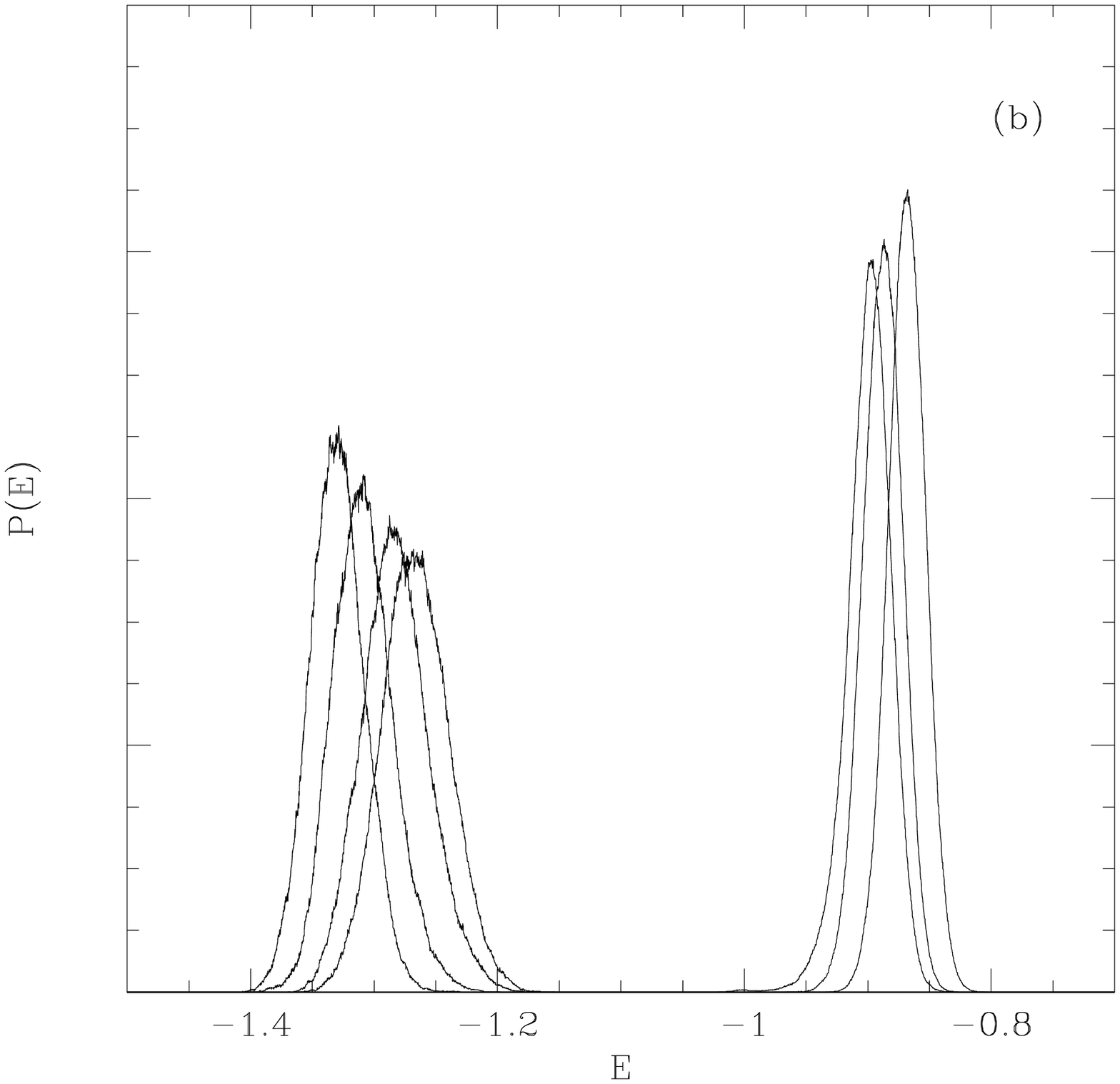}}\end{center}
\caption[99]{Monte-Carlo results for (a) the total energy per spin $E$
as a  function of $T$ (open  and  filled  circles correspond to heating
and cooling runs, respectively);  (b) energy histograms around $T=1.5$ (full
line in (a)) for a frustration parameter $Q=0.4$.}  \end{figure}
\begin{figure}
\begin{center}
 \resizebox{10.6cm}{4.8cm}{\includegraphics{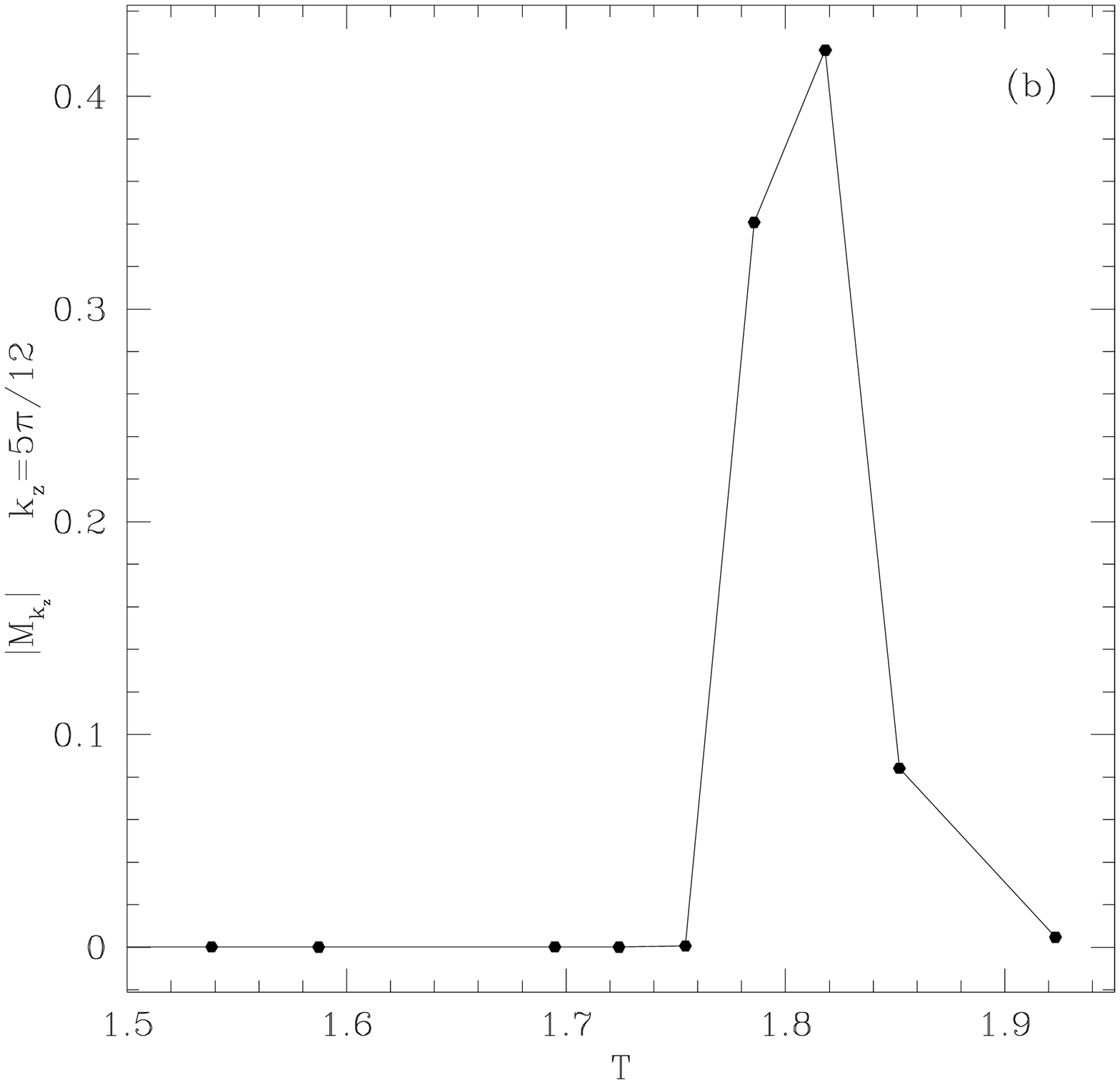}
\includegraphics{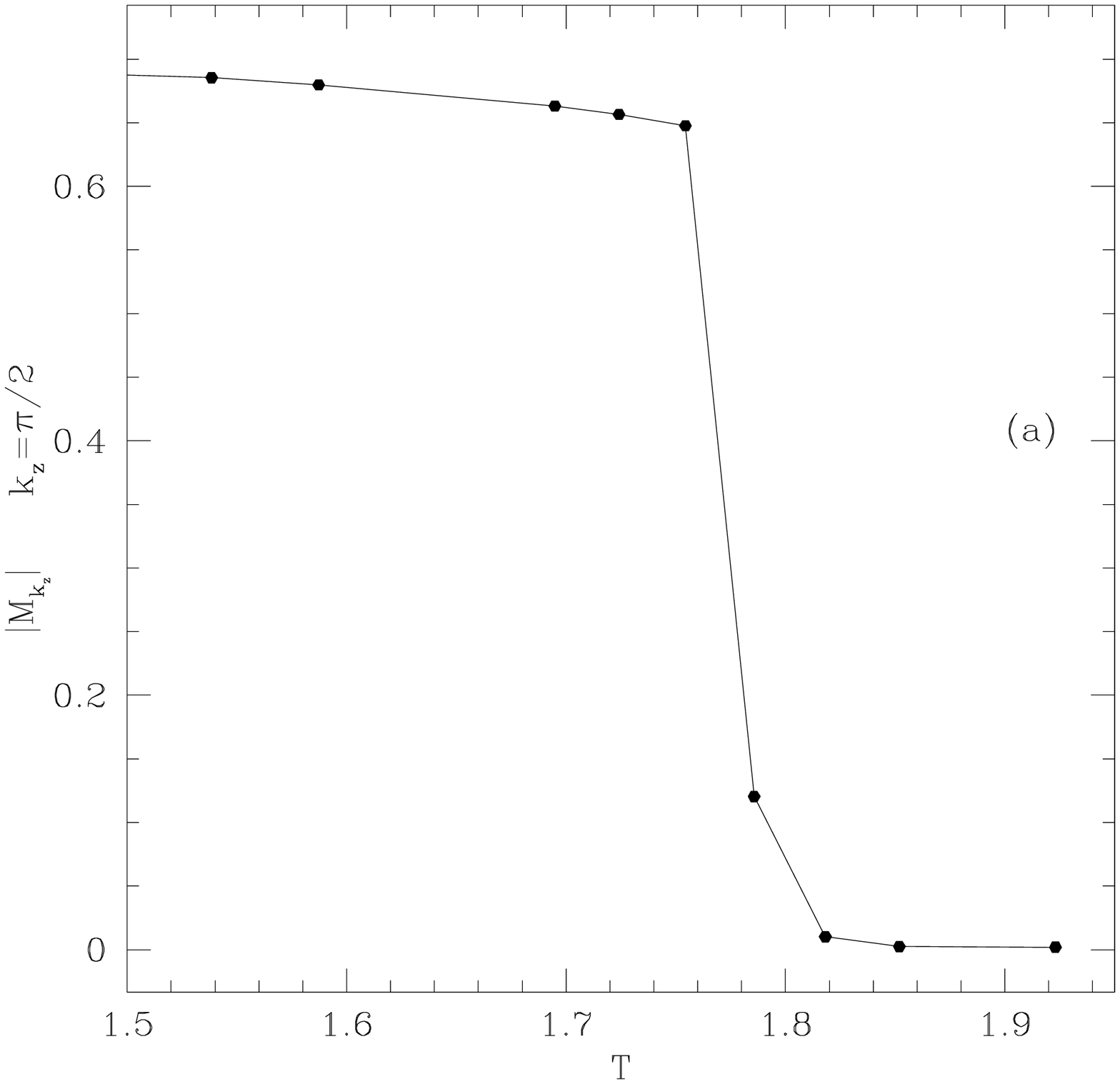}}
\end{center}
\caption[99]{Variation with $T$  of    the $|M_{\bf k}|$    with ${\bf
k}=(0,0,k_z)$ for (a)  $k_z=\pi/2$  and  (b) $k_z=5\pi/12$  from   the
Monte-Carlo  simulations  (for  $Q=0.144$  and a  box  size =$12\times
12\times  24$).    The microphase   separation  transition occurs  for
$T_0\simeq 1.84$ and it  is signaled by  the  appearance of a  nonzero
value of $|M_{k_z}|$     for $k_z=5\pi/12$.  At a  lower   temperature
$T\simeq    1.77$, a second  transition   occurs  to a lamellar  phase
characterized  by a  nonzero   $|M_{k_z}|$ for $k_z=\pi/2$ (all  other
$M_{{\bf k}}$'s are zero).}  \end{figure}

In  this  letter, we have  studied   the properties  of the microphase
separation transition and of the modulated  phases observed in a three
dimensional Coulomb  frustrated Ising model.  Our  Monte-Carlo results
provide strong evidence  that the   transition from the   paramagnetic
phase is  a  fluctuation-induced first-order one,  in  contrast to the
Ising ferromagnet with  dipolar interactions for which  the transition
is second-order\cite{binder}.   The only (second-order) critical point
in the temperature-frustration phase diagram is at $T^0_c$.  Since the
transition line approaches $T^0_c$ from below in a nonanalytic fashion
as $Q^{\kappa}$ with $\kappa\sim 0.25-0.35$, avoided critical behavior
can still dominate  the physics of weakly  frustrated system above the
transition  line, thereby    leading   to frustration-limited   domain
structures   in this region.    Additional  simulation studies of  the
paramagnetic     state  will however  be    necessary  to confirm this
prediction.  We have also shown that the model gives rise to a variety
of  modulated phases whose characteristics  vary both with temperature
and frustration.
The Laboratoire de Physique Th\'eorique des Liquides is UMR No 7600 au
CNRS.  We  are  grateful to D.  Kivelson, S.   Kivelson, Z.  Nussinov,
A. Lesne, and J.-M. Victor for many stimulating discussions
\begin{figure}
\begin{center}
\resizebox{14.7cm}{4.7cm}{\includegraphics{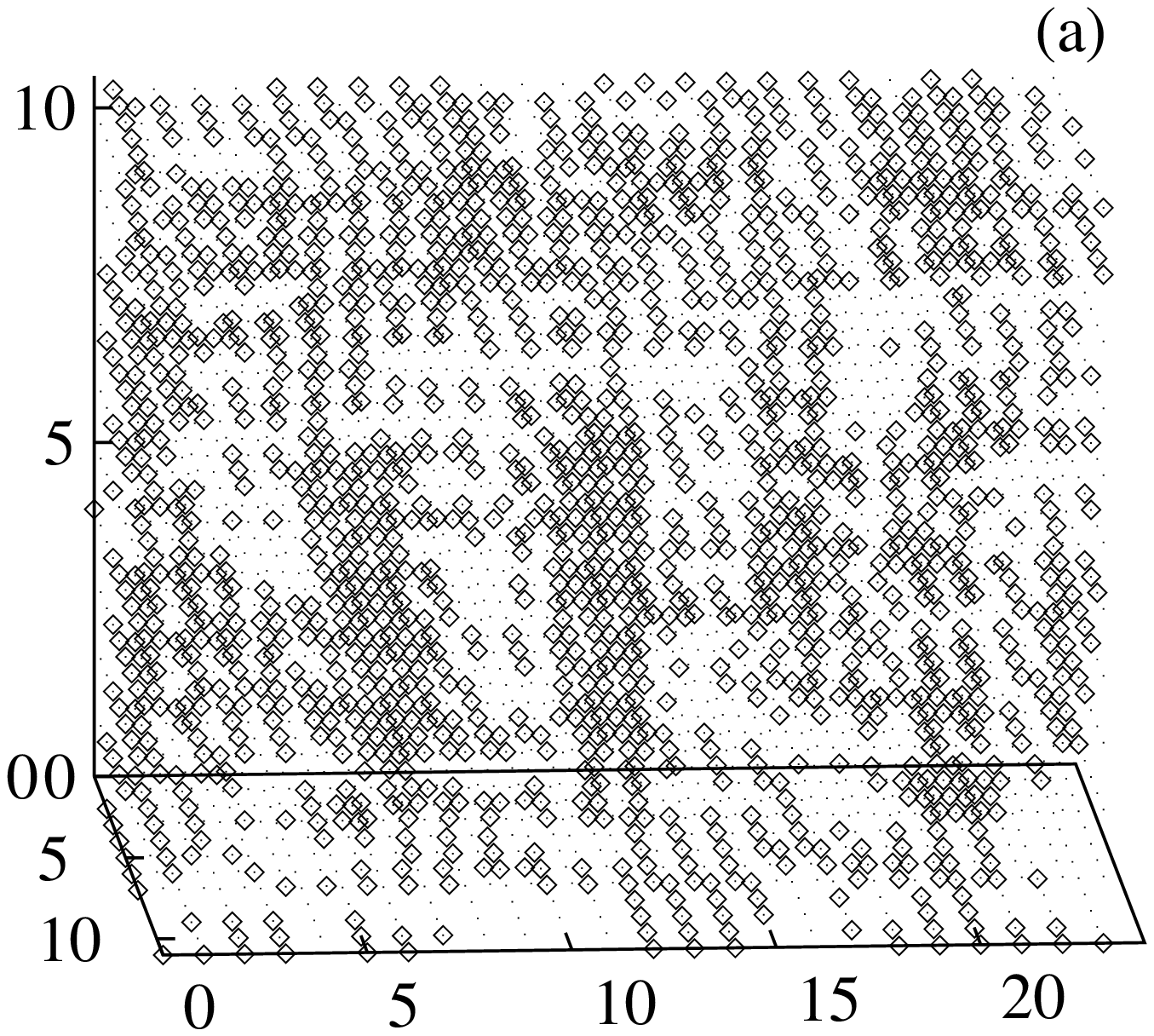}
\includegraphics{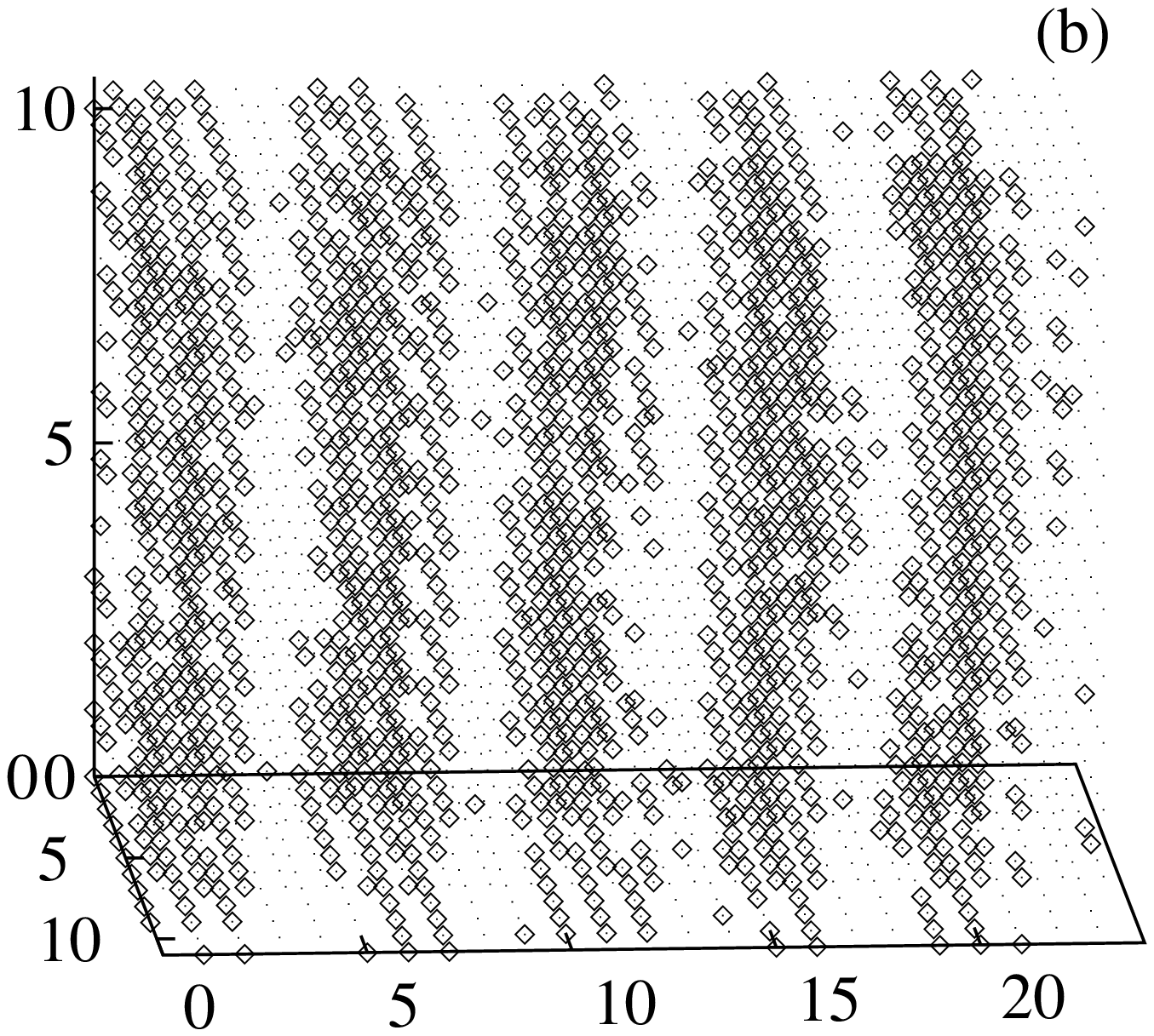}\includegraphics{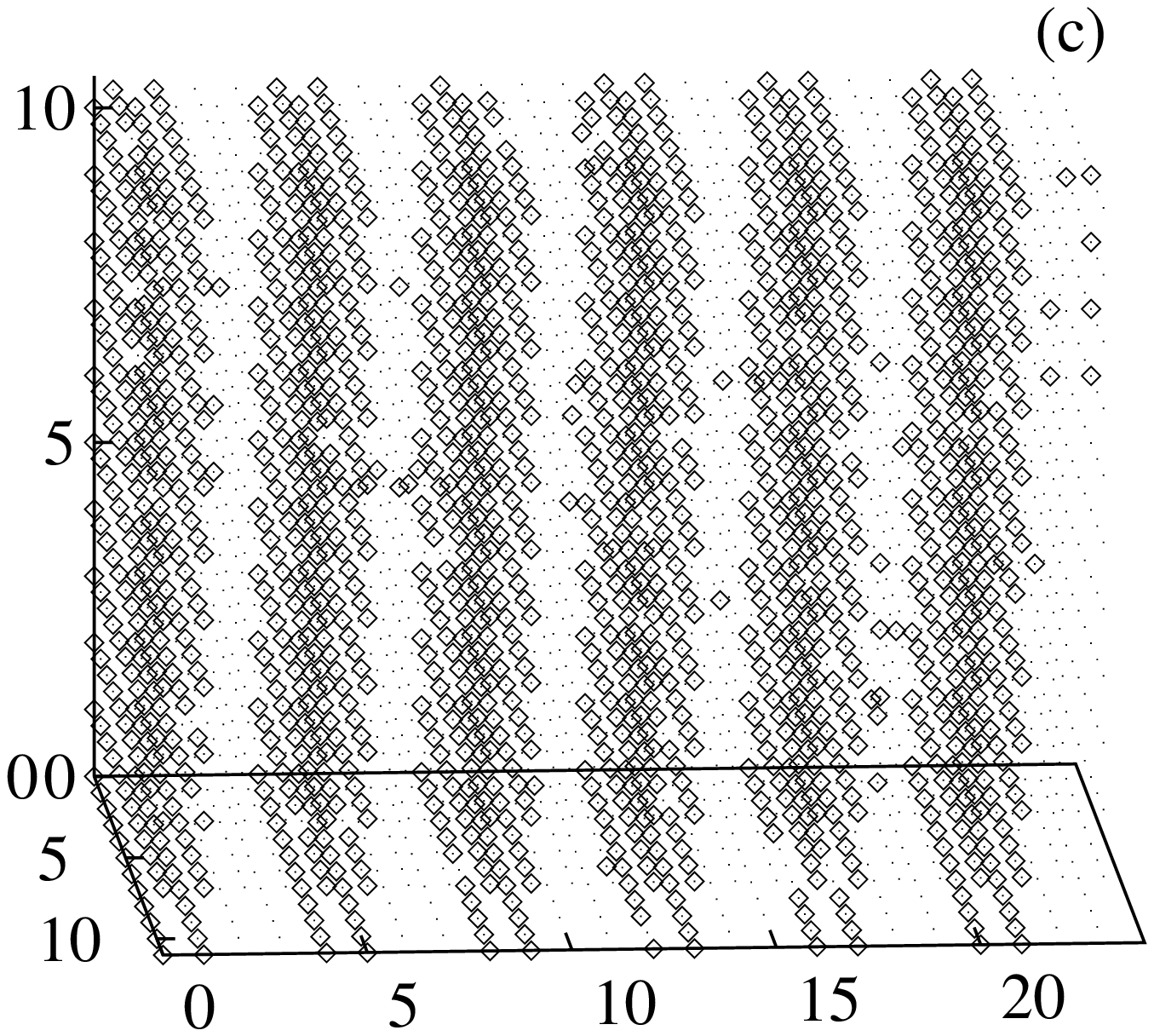}}
\end{center}
\caption[99]{Spin configurations corresponding to the different phases
observed in the Monte-Carlo simulations at $Q=0.144$  (box size $12\times 12\times
24$):  (a) disordered (paramagnetic)  phase, (b)  modulated phase with
averaged half-period of 2.4 lattice units; (c) lamellar structure with
 a half-period of 2 (this is also the ground  state). Up and down
spins are represented by diamonds and dots, respectively.} 
\end{figure}


\begin{thebibliography}{99}
\bibitem{leibler} L.   Leibler, Macromol.   {\bf 13}, 1602  (1980); T.
Ohta and K.     Kawazaki,  Macromolecules  {\bf 19},   2621    (1986);
G.H.  Fredrickson, and  E.  Helfand, J.   Chem.  Phys.   {\bf 87}, 697
(1987). 
\bibitem{bates}   F.S.     Bates         and  G.H.        Fredrickson,
Ann. Rev. Phys. Chem {\bf 41}, 525 (1990); M.W. Matsen and F.S. Bates,
Macromol. {\bf 29}, 1091 (1996). 
\bibitem{fried} H. Fried and K. Binder, J. Chem.  Phys. {\bf 94}, 8349
(1991);  M. Schulz  and K.  Binder,  J.  Chem.   Phys.  {\bf 98},  655
(1993). 
\bibitem{WCS} D. Wu, D.  Chandler, and B.  Smit, J.  Phys. Chem.  {\bf
 96}, 4077 (1992).; M.W. Deem and D.  Chandler, Phys. Rev.  E{\bf 49},
 4268 (1994). 
\bibitem{carraro} C. Carraro, Physica. A{\bf 236}, 130 (1997). 
\bibitem{KKZNT} D. Kivelson, S.A.  Kivelson, X. Zhao, Z. Nussinov, and
 G. Tarjus, Physica A{\bf 219}, 27 (1995). 
\bibitem{EK}  V.J. Emery and  S.A.  Kivelson, Physica  C{\bf 209}, 597
(1993). 
\bibitem{SH} J. Swift and P.C Hohenberg,  Phys.  Rev.  A {\bf 15}, 319
(1977); C.  Roland and R. Desai, Phys.   Rev. B {\bf 42}, 6648 (1990);
K.R. Elder, J. Vi\~nals and M. Grant, Phys. Rev.  Lett, {\bf 68}, 3024
(1992); S. Glotzer, D.  Stauffer, and N.  Jan, Phys.  Rev. Lett.  {\bf
72}, 4109 (1994). 
\bibitem{GTV} M. Grousson, P. Viot and G. Tarjus, in preparation. 
\bibitem{CEKNT} L. Chayes, V.J.  Emery, S.A Kivelson, Z. Nussinov, and
 G. Tarjus, Physica A{\bf 225},  129 (1996); Z. Nussinov, J.  Rudnick,
 S.A.  Kivelson, and L. Chayes, preprint. 
\bibitem{B} S.A. Brazovskii, Sov. Phys. JETP {\bf 41}, 85 (1975). 
\bibitem{MH} D. Mukamel, and R.M. Hornreich.  J.  Phys. C{\bf 13}, 161
(1980);  D.  Ling,  B.  Friman,  and G.  Grinstein,  Phys.  Rev. B{\bf
24}, 2718 (1981). 
\bibitem{LEFK} U. L\"ow, V.J. Emery, K.  Fabricius, and S.A. Kivelson,
Phys. Rev. Lett. {\bf 72}, 1918 (1994). 
\bibitem{mcisaac} A.B.  MacIsaac, J.P.  Whitehead, M.C.  Robinson, and
 K. De'Bell, Phys. Rev. B{\bf 51}, 16033 (1995).
\bibitem{binder} R. Kretschmer and K. Binder, Z. Physik B{\bf 34}, 375
(1979). The effect of the dipolar interaction is less drastic that of the
coulombic interaction studied here since complete ferromagnetic
ordering is not prevented and the order of the transition is unchanged.

\bibitem{selke} W. Selke, Physics Reports {\bf  170}, 213 (1988). Note
 that a   major difference with the   ANNNI model is   that  complete
 ferromagnetic order is forbidden for all nonzero values of the
 frustration parameter $Q$, so that the Coulomb frustrated Ising model
 does not have an ordinary Lifshitz point.  In  addition, the  T=0 phase
 diagram already displays a rich variety of modulated phases. 
\end{thebibliography}
\end{document}